# Spin-torque memristors based on perpendicular magnetic tunnel junctions with a hybrid chiral texture

Xueying Zhang[a,b,1], Wenlong Cai[a,1], Mengxing Wang[a,1], Kaihua Cao[a], Tianrui Zhang[a], Houyi Cheng[a,b], Shaoxin Li[a,b], Daoqian Zhu[a], Weisheng Zhao[a,b,2]

[a]*Fert Beijing Institute, BDBC, School of Microelectronics, Beihang University, Beijing 100191, China*

[b]*Beihang-Goertek Joint Microelectronics Institute, Qingdao Research Institute, Beihang University, Qingdao 266000, China*

[1]These authors contributed equally to this work

[2]To whom correspondence should be addressed. Email: weisheng.zhao@buaa.edu.cn

## Abstract

Spin-torque memristors were proposed in 2009, which could provide fast, low-power and infinite memristive behavior for large-density non-volatile memory and neuromorphic computing. However, the strict requirements of combining high magnetoresistance, stable intermediate states and spin-polarized current switching in a single device pose difficulties in physical implementation. Here, we experimentally demonstrate a nanoscale spin-torque memristor based on a perpendicular-anisotropy magnetic tunnel junction with a CoFeB/W/CoFeB composite free layer structure. Its tunneling magnetoresistance is higher than 200%, and memristive behavior can be realized by spin-transfer torque switching. Memristive states are maintained by robust domain wall pinning around clusters of W atoms, where nanoscale vertical chiral spin textures could be formed through the competition between opposing Dzyaloshinskii-Moriya interactions and the fluctuating interlayer coupling caused by the Ruderman–Kittel–Kasuya–Yosida interaction between the two CoFeB free layers. Spike-timing-dependent plasticity is also demonstrated in this device.

**Keywords:** spintronics | memristor | magnetic tunnel junction | chiral spin texture | neuromorphic computing

## Significance

In this research, a nanoscale all-spin-torque memristor based on perpendicular-anisotropy magnetic tunnel junction with W-inserted dual free layers is experimentally demonstrated and the spike-timing-dependent plasticity functionality is validated in a single spintronic device. In addition, we demonstrate that vertical chiral vortices could form under the energy competition between the Ruderman–Kittel–Kasuya–Yosida interaction and the opposing interfacial Dzyaloshinskii-Moriya interactions in ferromagnet/heavy metal/ferromagnet structure. This new spin texture is shown to be the origin of the memristive behavior by creating robust domain wall pinning effect in the free layer. This work paves a way for the practical application of spintronic device on neuromorphic computation, which shows great potential to overcome the Von Neumann bottleneck.

Memristors are considered to be essential elements for realizing neuromorphic computing(1–3). Traditional memristors rely on ion motion and ionic valence changes in materials(1, 4, 5). However, most of them suffer from certain limitations, such as finite endurance or relatively low switching speed(1). Spin-torque memristors, in which the state is modulated by a spin-polarized current, provide an alternative solution(6). The concept of a spin-torque memristor, in which the domain wall (DW) motion in the free layer (FL) of a magnetic tunnel junction (MTJ) is used to obtain a memristive magnetoresistance, was first proposed in 2009(7). Nevertheless, no real device with all-spin-torque operation has been experimentally reported until now. The intermediate tunneling magnetoresistance (TMR) is difficult to stabilize against thermal activation or stimulus currents, especially in devices with nanoscale dimensions(8). An FL in the partially switched state with DWs has usually higher energy than monodomain states because both Heisenberg exchanges and magnetic anisotropy favor a collinear spin texture. Several possible solutions have been proposed, such as creating an intermediate state with the assistance of shape anisotropy(9) or manipulating memristive switching by means of DW pinning in some complex geometries(10, 11) or by engineering the reference layer(12). However, these solutions require a large device size or an external magnetic field for the practical realization of memristive behaviors. The interfacial Dzyaloshinskii-Moriya interaction (DMI)(13, 14), a form of antisymmetric exchange that favors a chiral spin texture, makes it possible to obtain intrinsically stable noncollinear magnetic structures in nanometer-scale magnet and provides new means to realize memristive MTJs with nanoscale dimensions(15, 16).

In this work, we experimentally demonstrate a nanoscale spin-torque memristor based on a perpendicular-anisotropy MTJ with W-inserted dual FLs. A high TMR ratio, a low resistance-area product (RA), and spin-polarized-current-induced switching are achieved. The memristive behavior originates from robust DW pinning, which could arise from the fluctuations of the DW surface energy around clusters of W atoms under the competition between the Ruderman–Kittel–Kasuya–Yosida (RKKY) interaction and the interfacial DMIs(13, 14). The spike-timing-dependent plasticity (STDP) functionality is also validated based on this spin-torque memristor.

## Results

### Dual-FL MTJ device

Figure 1*A* introduces the layer structure used to fabricate the memristive MTJ: synthetic antiferromagnet (SAF)/W (0.25)/CoFeB (1.0)/MgO (0.8)/CoFeB (1.3)/W (0.2)/CoFeB (0.5)/MgO

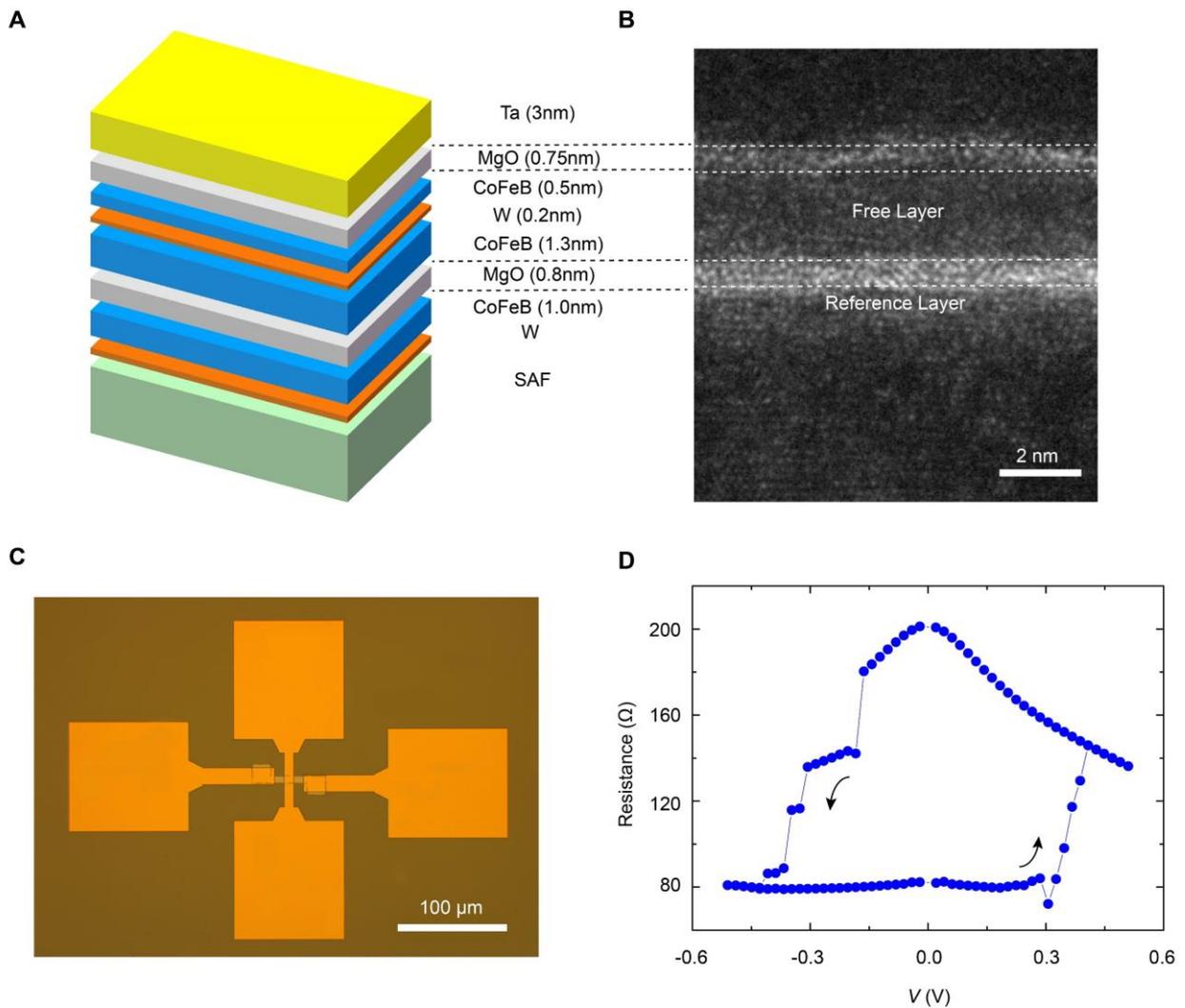

**Fig. 1. Layer structure and electric test of the device.** (*A*) The stack structure of the MTJ. (*B*) Cross-sectional TEM image showing the layer quality of the stack. (*C*) Optical microscopy image of an MTJ with four electrodes. (*D*) STT-induced switching as measured by means of a long voltage pulse with $\tau$ = 100 ms. A 20-mT perpendicular field was constantly applied during the test to compensate the bias of stray field from the reference layer, the same is also true throughout the manuscript.

(0.75)/Ta (3.0). The stack was prepared with a Singulus magnetron sputtering machine and was annealed at 390°C for 1 hour after deposition. In contrast to traditional MTJs with a single-FL structure, in this study, an atomic-thickness W layer was inserted between two FLs during sputtering deposition to engineer the FL properties(17, 18). A transmission electron microscopy (TEM) image of the multilayer stack is presented in Fig. 1*B*.

The dual-FL MTJ film was patterned into circular nanopillars with a 200-nm radius (R) using electron beam (e-beam) lithography and Ar ion milling and was instrumented with gold electrodes, as shown in Fig. 1*C*. Then, spin-transfer torque (STT)-induced magnetization switching was demonstrated using a voltage pulse with a duration ($\tau$) of 100 ms, presenting multiple intermediate states (Fig. 1*D*). A TMR ratio as large as 200% and an RA of 7 $\Omega$ $\mu m^2$ were obtained at room temperature. The switching current was on the order of MA $cm^{-2}$.

**Memristive tunneling magnetoresistance**

In order to demonstrate the stability of these intermediate states, we measured the resistance by applying a small reading voltage of 0.01 V after each STT-switching pulse, as shown in Fig. 2*A*. The small reading voltage allows to avoid the voltage-dependent effect of the antiparallel resistance (19), and the intermediate states showed correspondence with those in Fig. 1*D*, confirming the good stability of intermediate states. In addition, another STT switching measurement was performed using trains of shorter voltage pulses with $\tau$=200 ns, as shown in Fig. 2*B*. During this measurement, far more intermediate states appeared compared with the measurement shown in Fig. 1*D*. In addition, as can be observed from the minor loops, the intermediate states are stable even when the applied voltage is reversed. Figure 3 gives the current-voltage loop obtained by a train of $\tau$=200 ns switching voltage pulses. One can find that once the applied voltage increases over a threshold value, the resistance of the device varies quasi-continuously, which is typical memristive behaviors.

We also checked the stability of the intermediate states against external magnetic fields. First, the resistance-magnetic field hysteresis loop in a device with R=200 nm was measured at a low temperature, as shown in Fig. 2D. As expected, the coercive field is very large in such a small device and no intermediate state was observed. On the other hand, after creating an intermediate state with a STT current and then applying an external field, we found that the intermediate state remained stable under a relatively weak external magnetic field. The field needed to destroy the intermediate state was approximately 15.5 mT after offsetting the bias caused by the stray field from the reference layer.

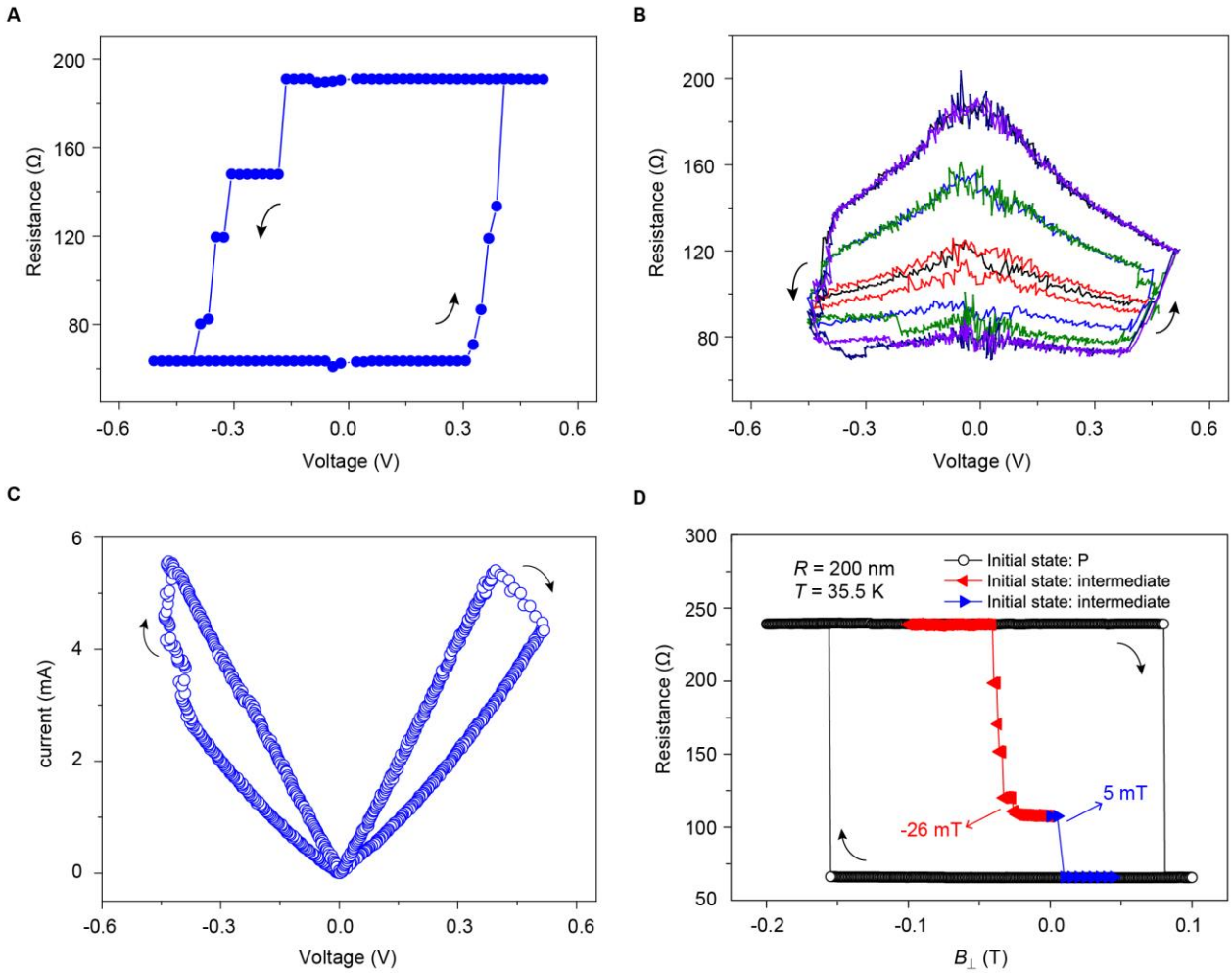

**Fig. 2. Spin-torque-induced and magnetic-field-induced switching of an MTJ with R=200 nm.** (*A*) Resistance-voltage loop. In this measurement, a train of voltage pulses with a duration of 100 ms and an increase of 0.02 V per step was applied. The resistance was measured using a voltage of 0.01 V after each stimulus pulse. (*B*) Resistance-voltage minor loops. In this measurement, short voltage pulses with $\tau$=200 ns were applied. (*C*) Current-voltage loop obtained using a train of voltage pulses with $\tau$=200 ns. The resistance was measured when switching voltage was applied in both *B* and *C*. (*D*) Black: full resistance-perpendicular field hysteresis loop of the device; red and blue: stability of intermediate state against external magnetic field at 35.5K.

## Synaptic plasticity

Due to its two-terminal nanoscale structure, tunable resistance, and non-volatile operation, this spin-torque memristor could be used for neuromorphic computing as synaptic devices(20). The plasticity, an essential property of an electronic synapse(21), was investigated by applying two different types of spike stimulation. Figure 3*A* shows the evolution of the resistance during the potentiation and depression cycles induced by a series of voltage pulses with $\tau$ = 200 ns. The magnitude of the spike voltage was increased in a stepwise manner, as shown in the top panel of Fig. 3*A*. When the applied pulsed voltage was above a certain threshold value (0.39 V/-0.37 V), the resistance of the device began to change gradually. After each spike pulse, the resistance was measured by a small voltage pulse (0.05 V and 1 μs), as shown in the inset of Fig. 3*A*. This non-volatile quasi-continuous variation is evidence of long-term functional synaptic plasticity(2). In the second case, the plasticity was explored by means of a train of identical voltage pulses with $τ = 200$ ns, as shown in Fig. 3*B*. The spike amplitude was chosen to be slightly larger than the threshold value to achieve analog resistance variation.

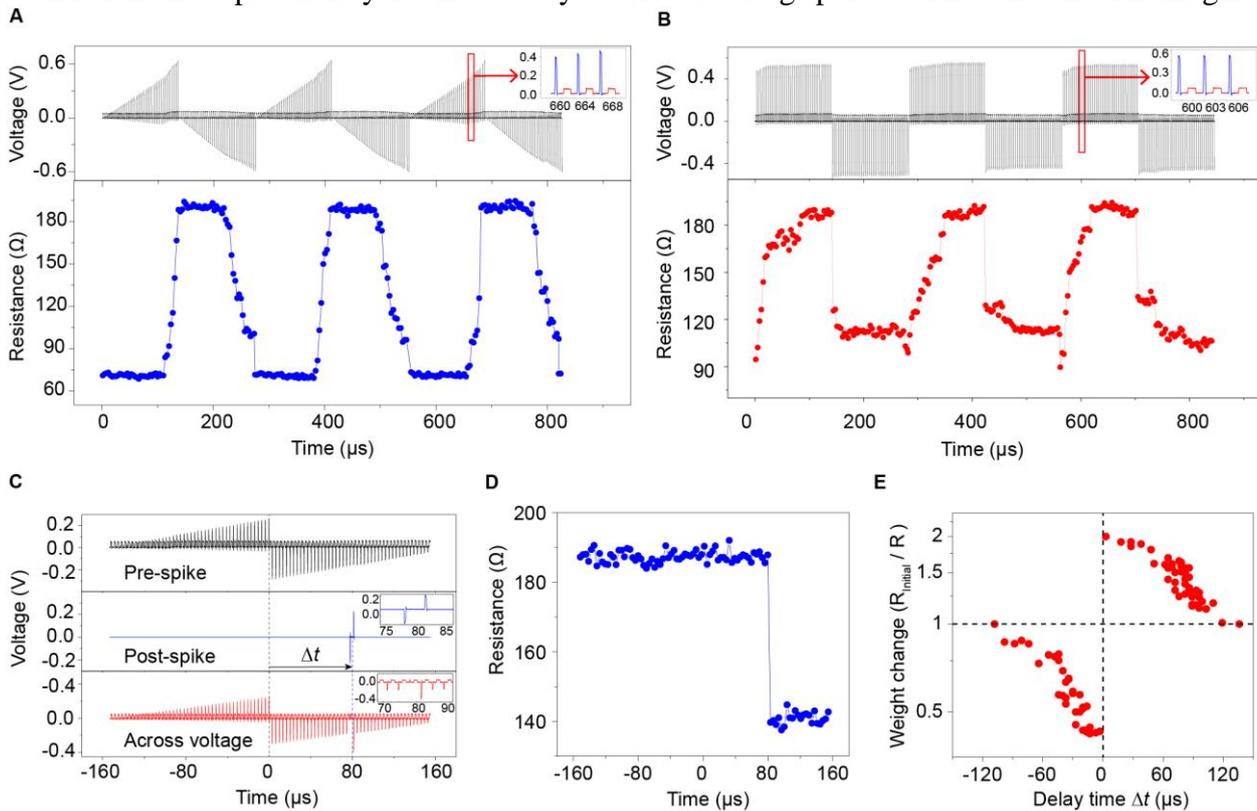

**Fig. 3. Synaptic plasticity.** (*A*) Plasticity explored by means of a ramped train of pulses. The upper plot shows the applied voltage pulses: a series of pulses with a $\tau$ = 200 ns duration and an increase of 0.01 V per step from 0 V to 6.2 V/-5.5 V was applied as the stimulus signal (shown in blue in the inset). A voltage pulse (0.05 V) with $\tau$ = 1 μs (red in the inset) was applied after each stimulus pulse as the detection signal. The lower plot shows the corresponding detected resistance. (*B*) Plasticity explored by means of identical pulses (0.54 V/-0.44 V) with $\tau$ = 200 ns. The resistance was characterized using a low-voltage pulse (0.05 V) with $\tau$ = 1 μs. (*C*) The pre-spike and post-spike waveforms and the across voltage obtained as the superposition of the pre- and post-spikes. (*D*) An example of the resistance variation corresponding to an across voltage with $\Delta t$ = 80 μs. (*E*) STDP learning curve. The changes in the synaptic weight were measured with different $\Delta t$ values ranging from -110 μs to 140 μs.

Based on the plasticity of the synaptic device, STDP(22, 23), which serves a fundamental learning function for artificial neural networks, was investigated. The pre-spike was a triangular-shaped voltage pulse constructed of several short pulses ($τ = 200$ ns) with a maximum amplitude lower than the threshold value, as shown in Fig. 3*C*. The post-spike consisted of two opposite-voltage pulses ($τ = 200$ ns and ±0.2 V). The delay time ($Δt$) is defined as the time elapsed from zero (the center time of the pre-spike) to the center time of the post-spike. The value of $Δt$ is positive (negative) when the pre-spike stimulation is applied before (after) the post-spike. Figure 3*D* shows the resistance variation corresponding to an across voltage with $Δt = 80$ μs, which is shown in Fig. 3*C*. The change in the synaptic weight is defined as the ratio of the initial resistance before spiking to the resistance after spiking. The weight change was characterized under different $Δt$ values ranging from -110 μs to 140 μs, and the results are presented in Fig. 3*E*. When the pre-spike occurs before (after) the post-spike, the synaptic weight increases (decreases), and a reduced effect are observed with a larger (smaller) $Δt$.

## Discussion

### Robust DW pinning effect in the dual FLs

The magnetoresistance of an MTJ depends on the relative magnetic state of the FL with respect to the pinned layer (PL)(19). A stable intermediate state is rare in traditional single-FL MTJs with submicron dimensions because once magnetic switching begins, the nucleated DW moves forward immediately and leads to the complete switching(8). By contrast, in the dual-FL MTJ fabricated here, the existence of multiple intermediate states suggests that the magnetic switching of the dual FLs is not complete at once.

To obtain a deeper insight into the FL properties, we deposited a MgO/CoFeB/W/CoFeB/MgO film with the same structure as that in the MTJ stack (called FL film in the following). The deposition and annealing conditions remained the same. As seen from the hysteresis loop of this film (see Fig. 4*A*) and the MTJ stack (see the supplementary material 1), the dual FLs are globally ferromagnetically coupled and show good perpendicular magnetic anisotropy (PMA). An intermediate state of the MTJ should not be caused by inconsistent magnetization of the upper FL (UpFL) and the lower FL (LwFL). However, the field-induced magnetization reversal of the FLs appears to be gradual, indicating that the DW propagation field in this film is much larger than the DW nucleation field. Next, both the field-induced and current-induced DW motion in the FL film were explored using Kerr microscopy. Figure 4*B* shows a dendritic trace after DW motion induced by a magnetic field of 3.6 mT, which is much rougher than that for an ordinary single-layer CoFeB film with low pinning effect(24). The velocity of the field-induced DW motion was measured and it leaves the thermally activated creep regime (25) when the driving field reaches $\mu_0 H_C$=16.3 mT, as shown in Fig. 4*C*. This critical field (conventionally called intrinsic pinning field of DW) can be seen as an indicator of the DW pinning strength in a magnetic material, below which DW motion is not possible without the assistance of thermal activation(25). This value is consistent with the field (15.5 mT) required to destroy an intermediate state in our MTJ device at low temperature, confirming that the intermediate states are stabilized by the strong DW pinning effect of the film.

In addition, we prepared microwires based on MgO/CoFeB/W/CoFeB/MgO stacks and tried to observe STT-driven DW motion. As shown in Fig. 4*D*, after the creation of several DWs via a short magnetic field pulse, an in-plane current was applied. No DW motion was observed even when the current density reached $2.4 \times 10^{11}$ A/m$^2$. When the current was further increased, maze domains appeared, which was probably a result of demagnetization caused by Joule heating. This result further confirms the difficulty of magnetic switching via STT-induced DW motion in the FLs of our device.

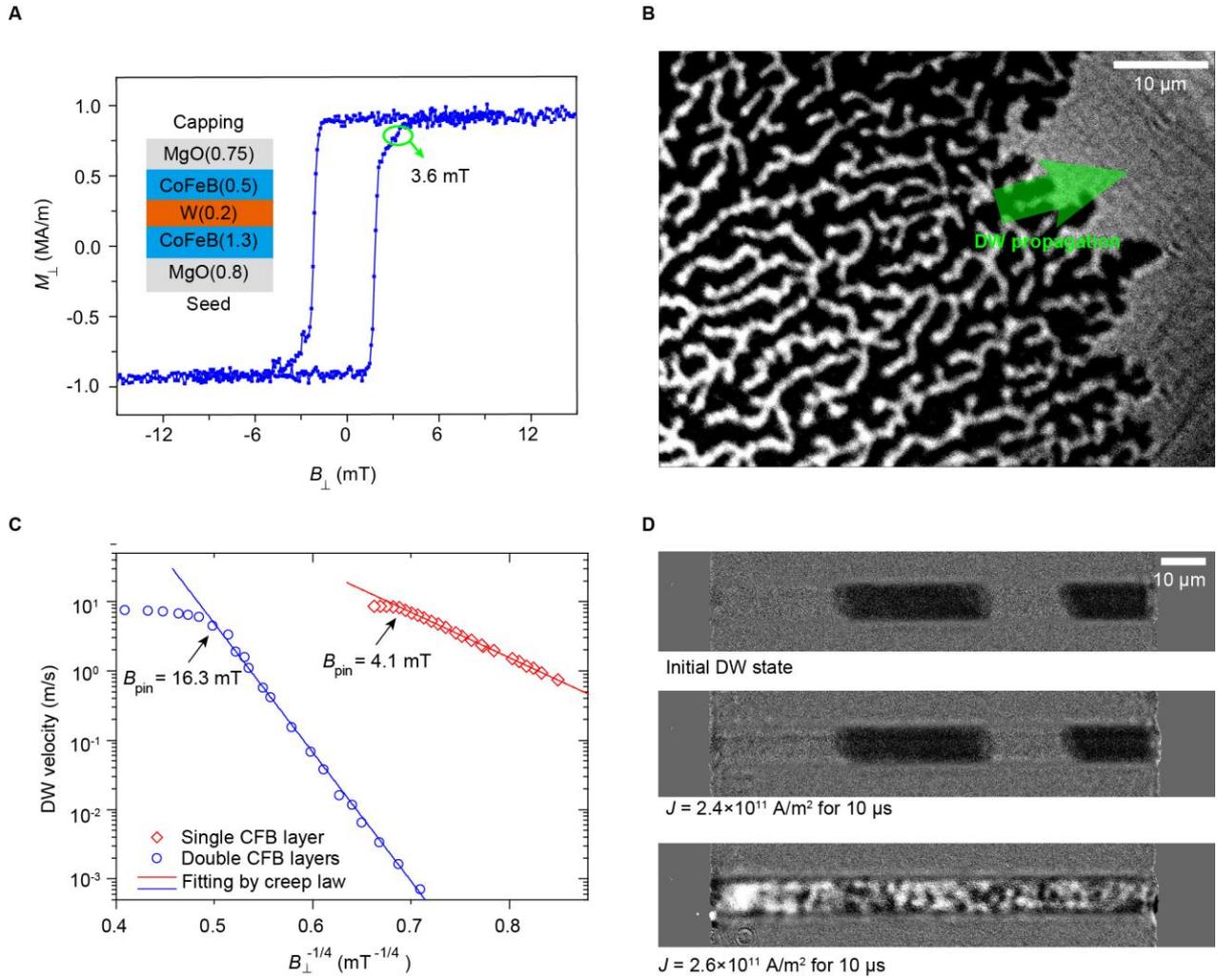

**Fig. 4. Strong DW pinning in a MgO/CoFeB/W/CoFeB/MgO film (FL film).** (*A*) The perpendicular hysteresis loop of the FL film. (*B*) Kerr image showing the dendritic trace of the domains after DW motion driven by a perpendicular field of 3.6 mT in the FL film. (*C*) Velocities of DW motion driven by a perpendicular field in FL film (in blue circle) and in a W/CoFeB(1.0 nm)/MgO single-magnetic layer film (in red diamond) and the linear fit using the creep law. (*D*) DW behavior driven by a current pulse in microwire patterned based on the FL film. The 1st image shows the magnetic state after DW creation using a field pulse, and the 2nd and 3rd images show DW states after the application of a current pulse.

**Hybrid DW structure induced by the interfacial DMIs and the RKKY interaction**

Conventionally, a heavy metal/CoFeB/MgO stack is very soft(24). Here, we measured the DW motion velocity in a W/CoFeB(1 nm)/MgO single ferromagnetic layer film for comparison and found the intrinsic pinning field was 4.1 mT, much lower than that for double FL films, as shown in Fig. 4*C*. In the dual FLs of our device, the pinning strength is increased by a factor of 4 after the insertion of the W layer. This strong pinning effect cannot be merely explained by DW energy fluctuations caused by the roughness of the interfaces since both the number of interfaces involved and the total thickness of CoFeB are doubled in the dual-FL film compared with the single-magnetic layer film(26).

As seen from the hysteresis loop, the 0.2-nm-thick W spacer results in a globally strong ferromagnetic exchange coupling between the UpFL and LwFL, consistent with the experimental measurements of S. Parkin(27). On the other hand, as seen from the energy-dispersive X-ray spectroscopy (EDS) mapping of our multilayer stack shown in Fig. 5*A*, the spatial distribution of W atoms is not homogeneous, with some stochastic overlap and some breakage. This inhomogeneous distribution of W may be caused by atom diffusion during the annealing process. According to the RKKY theory, two ferromagnets (FMs) separated by a thin metal layer can exhibit periodic ferro-/antiferromagnetic exchange, where the period depends on the type and lattice structure of the metal. Based on the experimental data reported by S. Parkin(27) and our fitting results according to the RKKY law, two FMs neighboring a W spacer begin to exhibit an antiferromagnetic coupling when the W thickness reaches approximately 0.44 nm, and the coupling reaches its peak at a W thickness of 0.55 nm, corresponding to approximately two atomic layers(28). Therefore, although the two FM layers in the FL structure of our stack exhibit global ferromagnetic exchange, the stochastic overlapping of the W atoms (W clusters) may lead to an antiferromagnetic exchange in some local regions.

On the other hand, considerable DMIs can arise at a W/CoFeB interface, because of the spin-orbit coupling(29, 30), and also at a CoFeB/MgO interface, which may be related to the charge transfer and electric field at this interface(31–33). This interaction is especially large for an Fe-rich CoFeB composition, as in our case, and promotes a chiral magnetic texture in the magnetic layer(34, 35). The chirality favored by the DMI depends on the materials and the direction of symmetry breaking. We have measured the DMI in films with different symmetry breaking via asymmetric DW motion when both the perpendicular field and in-plane field are applied. It is found that the DMI strength in a MgO/CoFeB(1.7 nm)/W film is about 0.45 mJ/m$^2$ and favors a left-handed chirality and the DMI strength in a W/CoFeB(1 nm)/MgO film is as large as 0.65 mJ/m$^2$ while favors a right-handed chirality (see supplementary materials S1). These results suggest that in the dual FLs of our MTJ, opposite chiralities are favored for the DWs in the UpFL and LwFL. On the one hand, because of the global ferromagnetic coupling of the two FLs, the total DMI cancels out (only 0.02 mJ/m² in the dual-FL film) and the DW configuration shows no obvious chirality (see the supplementary material S1). In contrast, when the exchange coupling between the UpFL and LwFL is weak or even antiferromagnetic in a local region where W atoms overlap, the DW chiralities in each FL should be determined by the corresponding DMIs. The magnetization in the DWs center in the two FLs are opposite, and a chiral vortex could form around the W cluster.

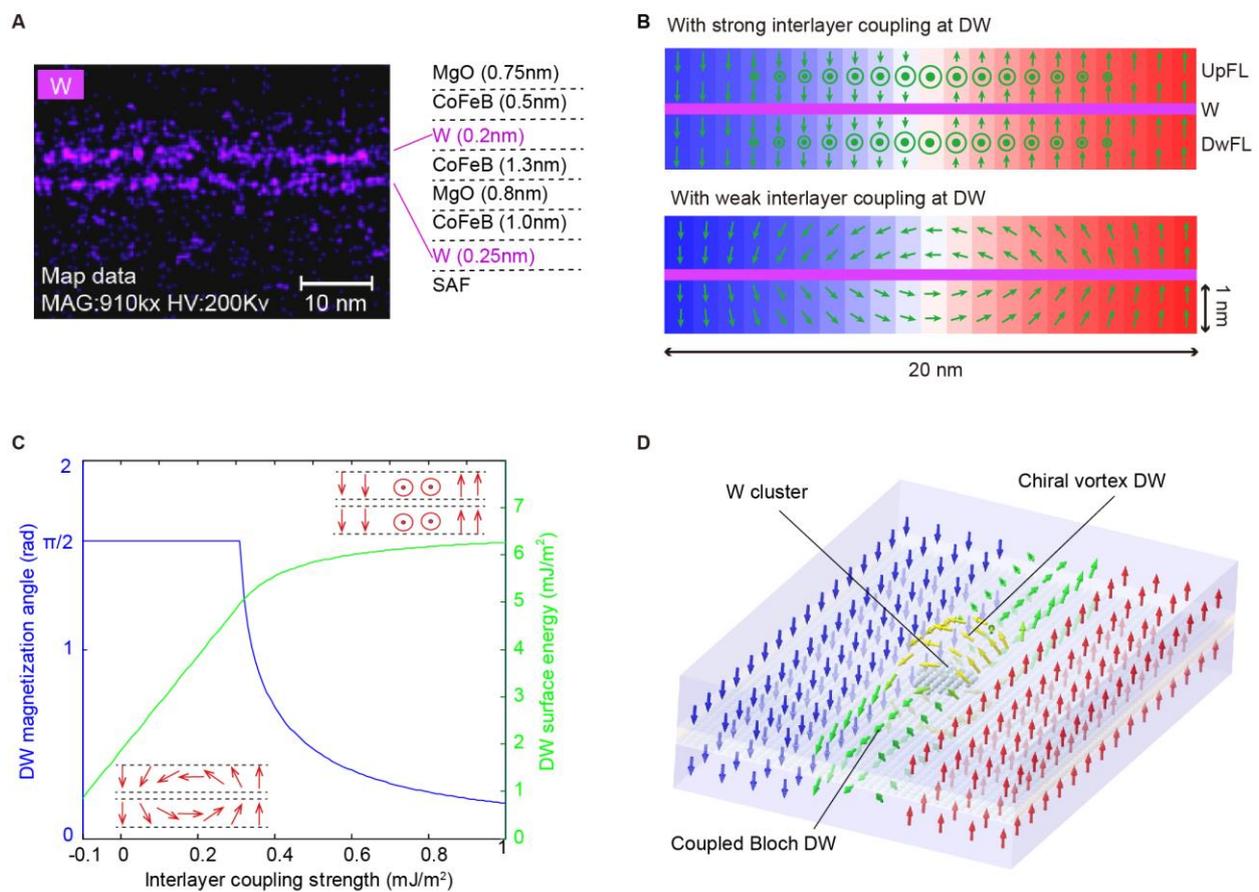

**Fig. 5. DW profile in CoFeB/W/CoFeB multilayers dominated by the competition between interlayer coupling and opposing DMIs.** (*A*) EDS mapping showing the inhomogeneous distribution of W atoms in the multilayer stack. (*B*) Side-view DW profiles under different interlayer coupling strengths obtained via micromagnetic simulations. (*C*) Variations in the azimuth angle of the DW magnetization and the surface energy as functions of the interlayer coupling strength. (*D*) Schematic showing the chiral vortex DW structure around a W cluster.

We simulated the magnetic states of two 1-nm-thick FM layers separated by a 0.2-nm spacer using the OOMMF code, as shown in Fig. 5B. In the first (second) case, the ferromagnetic coupling strength $J_{ex}$ was set to 1 mJ/m² (or 0.01 mJ/m²), where a positive (negative) sign of $J_{ex}$ means ferromagnetic (antiferromagnetic) coupling. The interfacial DMI constant D was set to 0.5 mJ/m² in both cases, with a negative (positive) sign in the UpFL (LwFL). The results confirm the above analysis: with strong ferromagnetic coupling, a Bloch DW configuration with parallel magnetization forms in the two FLs; with weak ferromagnetic coupling, a vertical vortex with a chirality dominated by the DMIs is formed.

We calculated the DW energy when interlayer exchange coupling and DMIs are involved. For simplicity, we suppose that the two FLs are totally mirror symmetric with respect to the spacer layer. The tilt angle of the magnetization in the center of a DW ($\vec{m}_{DW}$) with respect to the transverse direction is $\theta$ ($-\theta$) in the LwFL (UpFL), where $\theta \in [0, \pi/2]$. The total magnetic energy of the DW is(36, 37)

$$\sigma_{DW} = \sigma_0 + J_{ex}(1 - \cos 2\theta)\Delta/t_M - 2\pi D \sin\theta + \sigma_d \tag{1}$$

where $\sigma_0 = 4\sqrt{A_{ex}K_{eff}}$ is the surface energy of a Bloch DW with $\theta$=0(38), as described in the first case in Fig. 5B, without considering the DMI energy, $A_{ex}$ is the exchange stiffness, $K_{eff}$ is the effective anisotropy constant, $\Delta$ is the DW width, $t_M$ is the thickness of a single FL, and $\sigma_d$ is the energy associated to the dipole-dipole interaction. According to our numerical calculation and fitting, the last term can be approximately described by a simple expression $\sigma_d \approx A(1 + \cos 2\theta)$, where A≈0.78 mJ/m² in our case (39) (see the supplementary material S1). By minimizing the DW surface energy using $\frac{\partial \sigma_{DW}}{\partial \theta} = 0$, we obtain the value of θ at equilibrium:

$$\theta \approx \begin{cases} \arcsin \frac{\pi D t_M}{2J_{ex}\Delta - 2A*t_M} & for \left|\frac{\pi D t_M}{2J_{ex}\Delta - 2A*t_M}\right| < 1 \\ \frac{\pi}{2} & otherwise \end{cases} \quad (2)$$

For $K_{\text{eff}} = 1.25 \times 10^5 J/m^3$, as measured via vibrating sample magnetometry (VSM) for the FLs, $A_{\text{ex}} = 1.3 \times 10^{-11}$ J/m (40), $\Delta$=5 nm (37), $t_M$=1 nm, and $D$=±0.5 mJ/m², the variations of θ and the DW surface energy as functions of the interlayer coupling strength $J_{ex}$ are plotted in Fig. 5C. As the interlayer coupling strength decreases below 0.35 mJ/m², the DW profile transforms to a chiral vortex, and the DW surface energy drops rapidly. In fact, the formation of the chiral vortex DW minifies both the energy associated with DMIs and with dipole-dipole interactions (39) (see also the supplementary material S1).

According to our fitting results based on the data reported by S. Parkin(27), the interlayer coupling strength of two FM layers neighboring an intervening W layer decreases from 0.6 mJ/m² to -0.03 mJ/m² as the thickness of the W increases from 0.2 nm to 0.55 nm. In our dual-FL MTJ with a W spacer, when a moving DW encounters a W cluster with a dimension at the same order as the DW width, the DW configuration will locally transform into a chiral vortex, as illustrated in Fig. 5D. According to the above calculations, the surface energy of a chiral vortex DW configuration is several times smaller than that of a Bloch-type DW configuration. This energy difference serves as a deep energy well and will robustly pin the DWs (see the video showing this effect in supplementary material S2). Therefore, the energy wells induced by the balance between the DW configurations of the Bloch type and the chiral vortex type around W clusters should be the origin of the observed strong pinning effect.

Moreover, the dendritic domains after field-driven DW motion or after the application of a strong current also suggest that the strong demagnetizing field due to the thick FLs in this sample (1.8 nm in total) plays a non-negligible role. This demagnetizing effect also helps to maintain a multidomain state in the FLs (see the supplementary materials S1).

## Conclusion

In conclusion, a spin-torque memristor with a high TMR ratio, a low RA, a low working current, and a nanoscale size has been obtained by engineering an atomic-thickness W spacer between dual FLs in an MTJ. A memristive TMR has been achieved during STT-induced switching in both directions. By observing the field- and current-induced DW dynamics in the FL structure and comparing the stability of the intermediate TMR states at low temperature, we have demonstrated that the memristive behavior of the device originates from a robust DW pinning effect in the FLs. Furthermore, through theoretical calculations and micromagnetic simulations, we have proven that a chiral vortex DW can form around a cluster of W atoms because of the opposing interfacial DMIs and the RKKY interaction. The observed strong DW pinning could be caused by the energy well formed when the structure of a moving DW transfers between a coupled trivial Bloch configuration and a vertical chiral vortex configuration due to the inhomogeneous distribution of W atoms. The synaptic performance of the device has been studied systematically, and the STDP functionality has been validated. The realization of this nanoscale spin-torque memristor offers new opportunities for the application of spintronic devices in neuromorphic computing.

## Materials and Methods

1. Device preparation

The MTJ films were deposited by a Singulus TIMARIS 200 mm magnetron sputtering machine at a base pressure of $3.75 \times 10^{-9}$ Torr. Circular nanopillars were patterned by e-beam lithography, followed by Ar ion milling and SiO2 insulation. After the lift-off procedure, Ti/Au electrodes were evaporated for measurements.

2. Characterization of the film

   Fundamental properties of the MTJ stack and the FL stack, including the saturation magnetization and the perpendicular anisotropy, were both measured via VSM. In addition, the domain structure and DW motion in the FL film were observed using a Kerr microscope with a 400-nm resolution. A fast-perpendicular field with a rise time of sub-microsecond, which is produced by a small magnetic coil, is used to measure the DW motion velocities in the magnetic film. The strength of DMIs in films with different structure asymmetries are measured via analyzing DW motion velocities when out of plane magnetic fields and in-plane fields are applied simultaneously (see supplementary material S1).

3. Electrical measurements of the device

   The setup for the electrical characterization of the spin-torque memristor consists of a Lake Shore CRX-VF cryogenic probe station, Keithley 6221 current sources, and 2182 nanovolt meters.

4. Micromagnetic simulation

   Micromagnetic simulations based on the OOMMF code were performed to observe the magnetic texture and domain wall pinning effect in the dual FLs. Simulations based on Mumax3 software were performed to quantify the effect of the demagnetizing field on the domain wall structure. Both simulations are based on solving the Landau–Lifshitz–Gilbert equation.

## Acknowledgments

The authors thank the National Natural Science Foundation of China (Grant No. 61627813, 61571023), the International Collaboration Project B16001, the National Key Technology Program of China 2017ZX01032101, the VR innovation platform from Qingdao Science and Technology Commission, and Magnetic Sensor innovation platform from Laoshan District for their financial support of this work.

## Author contributions

W.Z. initialized, conceived and supervised the project; M.W. fabricated the devices, W.C., S.L. and W.Z. performed the measurements; X.Z. performed the simulation and calculation; X.Z., W.C., M.W and W.Z. wrote the manuscript; K.C. optimized the e-beam lithography flow and TEM characterization; H.C. optimized the film deposition technique; T. Z. and D.Z. analyzed the data. All authors discussed the results and implications.

## Competing interests

The authors declare no competing interests.

## Additional information

Two supplementary materials are available for this paper.

# Supplementary materials

**1. More information about the stack**

(1) Saturation magnetization and perpendicular anisotropy energy

Both the in-plane and out-of-plane (perpendicular) hysteresis loops of the film used to fabricate the magnetic tunnel junction (MTJ) were measured via vibrating sample magnetometry (VSM), as shown in Fig. S1A. Both the free layer (FL) structure and the reference layer exhibited good perpendicular magnetic anisotropy (PMA). Furthermore, the dual FLs were globally well ferromagnetically coupled.

After producing a film (FL film) with the same structure as the FL structure of the MTJ stack, we characterized the in-plane (Fig. S1B) and out-of-plane (Fig. 4A in the main text) hysteresis loops. The saturation magnetization $M_S$ of the FLs in the MTJ was estimated based on the perpendicular hysteresis loop of the FL film. $M_S$ was estimated to be approximately $1.0 \times 10^6$ A/m² at room temperature.

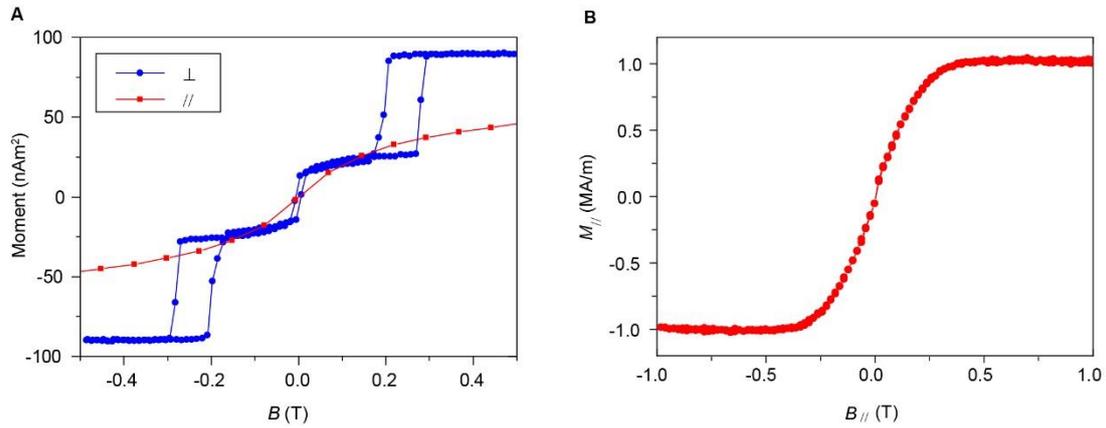

*Fig. S1. (A) Perpendicular and in-plane hysteresis loops of the multilayer stack used to fabricate the MTJ, measured via VSM at room temperature. (B), In-plane hysteresis loop of the FL film measured via VSM.*

The perpendicular anisotropy constant $K_U$ of the FLs was estimated based on the in-plane field hysteresis loop: the critical field that rotates the magnetization of the FLs to the in-plane direction is considered as the effective anisotropy field $\mu_0 H_{K,\text{eff}}$. The uniaxial anisotropy energy was calculated as follows(1): $K_U = \frac{1}{2}\mu_0 H_{K,eff} M_S + \frac{1}{2}\mu_0 M_S^2$. At room temperature, the effective anisotropy field $\mu_0 H_{K,\text{eff}}$ of the FL was measured to be 250 mT, and the anisotropy $K_U$ was estimated to be $7.5 \times 10^5$ J/m³.

(2) Dzyaloshinskii-Moriya interactions (DMIs) in films with different symmetry breaking

We have measured the DMI in two films with different symmetry breaking, i.e., a MgO/CoFeB(1.7 nm)/W film and a W/CoFeB(1 nm)/MgO film. At the same time, the overall DMI in the MgO/CoFeB/W/CoFeB/MgO dual FLs film is also measured. All films are annealed at the same temperature. Measurements are based on the asymmetric domain wall (DW) motion when driven by a perpendicular field under the application of a varying in-plane field(2, 3). High perpendicular fields $B_\perp$ are used so that the DW moves beyond the thermally activated creep regime. Pinning effects enhanced by the application of in-plane fields(4) have no influence in this regime and the measured DMI is believed to be more accurate(3). Results are shown in Figure S2. In each measurement, the perpendicular remains unchanged while the in-plane field varies step by step. The in-plane field under which the DW velocity reaches a minimum value can be regarded as the effective DMI field $H_{DM}$. Accordingly, the strength of the overall DMI can be estimated as $D = \mu_0 H_{DM} M_S \sqrt{A_{ex}/K_{eff}}$. Here, $A_{ex}$ is the Heisenberg exchange stiffness and $K_{eff}$ is the effective anisotropy constant.

It is found that the $\mu_0 H_{DM}$ in the W/CoFeB/MgO film is as large as 65 mT and the $\mu_0 H_{DM}$ in the MgO/CoFeB/W film is about 45 mT. The corresponding DMI constant is estimated to be 0.65 mJ/m$^2$ and 0.45 mJ/m$^2$. Here, $M_S$=1.0×10$^6$ A/m$^2$, $A_{ex} = 1.3 \times 10^{-11}$ J/m (5) and $K_{eff} = 1.25 \times 10^5 J/m^3$ is used. While the asymmetry of DW motion shows opposite direction, suggesting that the sign of the DMI in these two films is different. On the contrary, in the MgO/CoFeB/W/CoFeB/MgO dual FLs film, although some slight asymmetry of the DW motion velocities was observed, the $\mu_0 H_{DM}$ was found to be as small as 2 mT and the estimated DMI constant was calculated to be D=0.02 mJ/m$^2$. This is a very weak value. This weak DMI can be explained by the symmetry of the MgO\CpoFeB\W\CoFeB\MgO structure. Although a considerable DMIs exist at each MgO\CoFeB and CoFeB\W interface, the overall DMI cancels out.

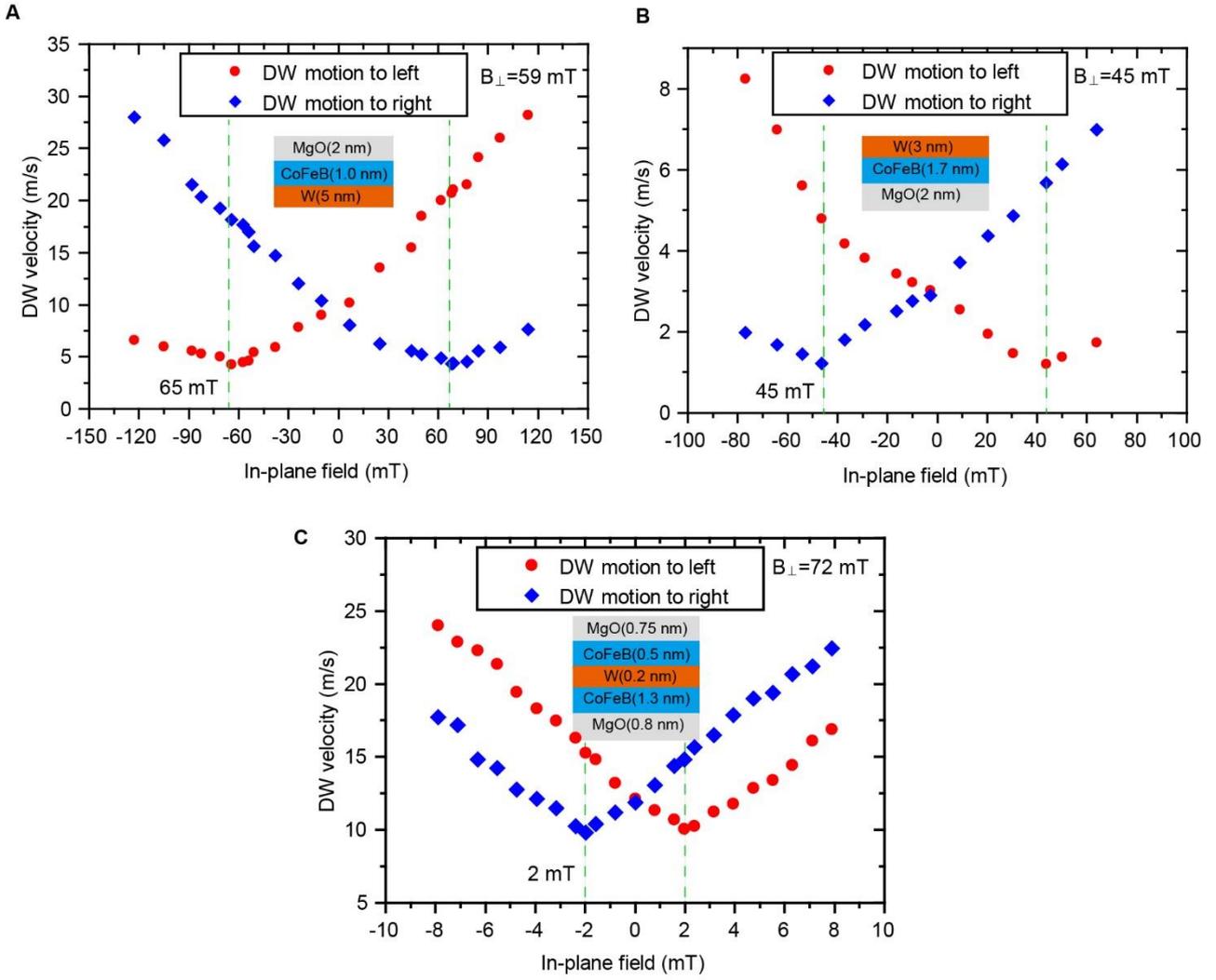

*Fig. S2. Measurements of the DMI in a three different film structures based on asymmetrical DW motion when an in-plane field is applied.* The film structure is: (**A**) W/CoFeB/MgO; (**B**) MgO/CoFeB/W; (**C**) MgO/CoFeB/W/CoFeB/MgO. The magnitude of perpendicular field $B_\perp$ used is given in upper right of each figure.

(3)     Ruderman–Kittel–Kasuya–Yosida (RKKY) exchange through the W spacer layer

S. Parkin measured the RKKY exchange of two magnetic layers separated by a W layer(6). The structure used was Si\Cr (3.5 nm)\[Co (1.5 nm)\W (x nm)]16\Cr (2 nm), with in-plane magnetic anisotropy in the Co layer. The following results were reported: the first antiferromagnetic coupling peak occurs at $t_W$=0.55 nm, and the antiferromagnetic coupling strength is 0.03 mJ/m2 at this peak. The thickness of W spacer that corresponds to this first antiferromagnetic region is approximately 0.3 nm.

We fitted these data using the well-known RKKY law(7):

$$J(t_W) = A \cdot F(2k_F t_W) \tag{S-1}$$

$$F(x) = \frac{x\cos x - \sin x}{x^4} \tag{S-2}$$

where *A* is a parameter related to the magnitude of the coupling strength and $k_F$ is a parameter related to the oscillation period. The fitting results are shown in Fig. S3. From this fit, we can conclude that the ferromagnetic coupling between the two magnetic layers is $J$ex=0.6 mJ/m² at $t_W$=0.2 nm and drops to zero at $t_W$=0.43 nm.

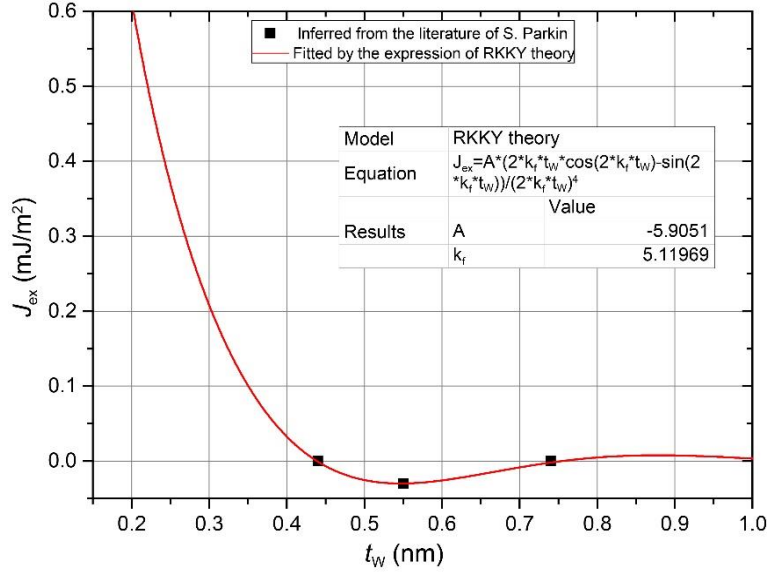

*Fig. S3. Strength of the RKKY coupling $J_{ex}$ of two ferromagnet (FM) layers as a function of the thickness of an intervening W spacer layer.* The black squares represent the data reported by S. Parkin [Phys. Rev. Lett. 67, 3598 (1991)], and the red line represents the fitting results based on the RKKY theory.

## 2.  Magnetic static energy for coupled Bloch and chiral vortex DW configurations

The local dipole-dipole interactions between the DWs in the two FLs promote the formation of a chiral vortex DW because this type of DW configuration has a lower magnetic static energy than a coupled Bloch DW(8). Since the theoretical calculation of this energy is very difficult, here, we perform a numerical calculation using the micromagnetic simulation software MuMax3(9).

The simulated structure is shown in Fig. S4. The thicknesses of the lower FL (LwFL), the W spacer and the upper FL (UpFL) were 1 nm, 0.2 nm and 1 nm, respectively. The area of the simulated zone was 32 nm×16 nm (In fact, 30 nm is the typical geometrical width of a DW in film with PMA(8)). The saturation magnetizations of the two magnetic layers were both $1.0\times10^6$ A/m, and the perpendicular anisotropy energy was $7.5\times10^5$ J/m³. The exchange stiffness in each layer was $A_{ex}=1.3\times10^{-11}$ J/m. A coupled Bloch DW configuration was created as the initial state. Then, the DW center magnetization $\vec{m}_{DW}$ was tuned by means of a locally applied external field. The angle of $\vec{m}_{DW}$ in the LwFL (UpFL) with respect to the y-axis was varied from 0 to π/2 (0 to -π/2). The total demagnetizing energy for the simulated system, $E_d$, was saved during the simulation.

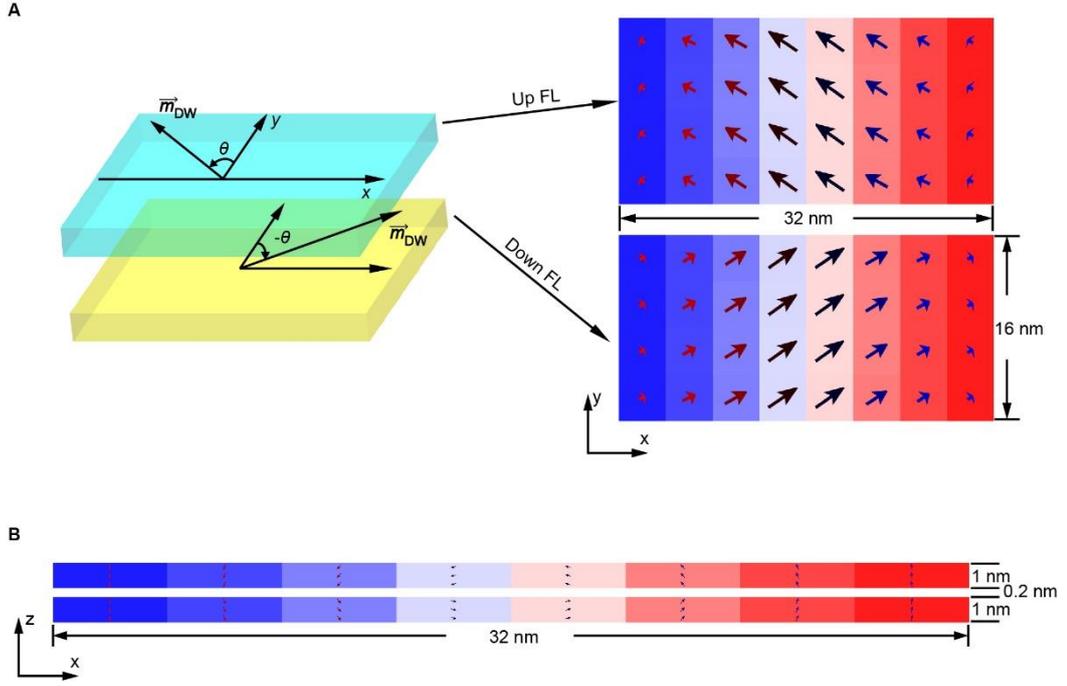

*Fig. S4. Estimation of the demagnetizing energy by means of a micromagnetic simulation.* **(A)** *Left: sketch of the simulated structure. Right: one frame showing the magnetizations of the UpFL and LwFL as viewed from the z direction during the simulation.* **(B)** *One frame showing the magnetic texture as viewed from the y direction during the simulation.*

Then, the DW energy per unit area associated with the dipole-dipole interactions was calculated as follows:

$$\sigma_d = \frac{E_d - E_{d0}}{t_m w} \quad \text{(S-3)}$$

where $E_{d0}$ is the demagnetizing energy of the whole simulated system when θ=π/2, which is taken as the reference level of the demagnetizing energy; $t_m$=1.8 nm is the total magnetic layer thickness; and w=16 nm is the width of the simulated structure, i.e., the length of the DW.

The variation in $\sigma_d$ as a function of $\theta$ is plotted in Fig. S5, which can be fitted by $\sigma_d \approx A(1 + \cos 2\theta)$. For this simulated configuration, the DW energy associated with the dipole-dipole interactions decreases by more than 1.6 mJ per unit area when the DW structure changes from the coupled Bloch configuration to the chiral vortex configuration. This is an important factor for the formation of chiral vortices.

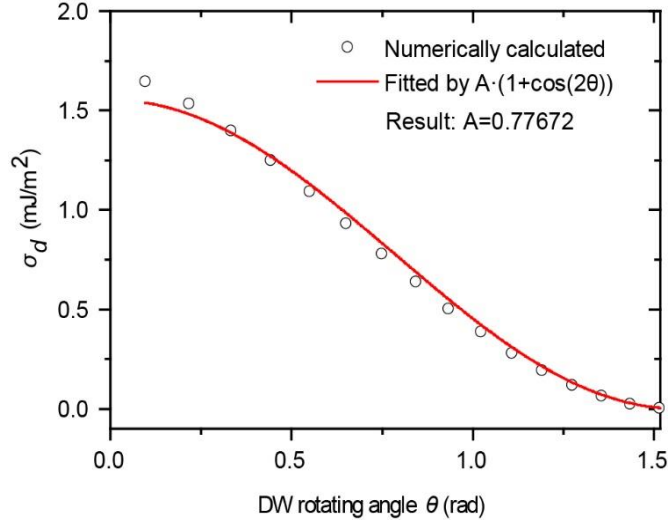

*Fig. S5. Variation of the DW energy in the two FLs contributed by dipole-dipole interactions as a function of the DW rotating angle θ, as extracted from micromagnetic simulations. The red line is the fitting results.*

3. **Effect of the demagnetizing field on the stability of an intermediate state**

The global demagnetizing energy of an FL is reduced if the FL is partially switched because the net magnetization approaches zero. In other words, a demagnetizing field $B_{demag}$ helps to stabilize a DW in an FL disc. Here, we use the concept of the effective magnetization current to evaluate the influence of such a demagnetizing field on the DW dynamics in the FL(10–12). For a uniformly magnetized film with perpendicular anisotropy, the demagnetizing field (or stray field) is equivalent to the Oersted field produced by a current loop around the edge of the film. The amplitude of this current is $I=M_S t_M$, where $M_S$ is the saturation magnetization and $t_M$ is the thickness of the magnet. Here, we suppose that the magnetic state of a partially switched FL is as shown in Fig. S6A. The DW is assumed to be a straight vertical line for simplicity of calculation, and the horizontal coordinate of the DW is $x_{DW}$. The $B_{demag}$ from the FL that is acting on the DW (at point P) is equal to the Oersted field from a current loop along the edge of the disc with the current direction shown in Fig. S6A. For $M_S=1\times10^6$ A/m, we numerically calculated the $B_{demag}$ acting on the DW as a function of the horizontal coordinate $x_{DW}$, and the results are shown in Fig. S6B. As seen from these results, the $B_{demag}$ always favors the stabilization of the DW in the center of the disc, and its magnitude increases as the DW approaches the edge. Moreover, the magnitude of $B_{demag}$ increases proportionally with the thickness. In our MTJ with a 1.8-nm-thick FL and a 200-nm radius, $B_{demag}$ increases to approximately 10 mT at the edge; thus, it plays a non-negligible role in stabilizing the intermediate state before complete switching is achieved.

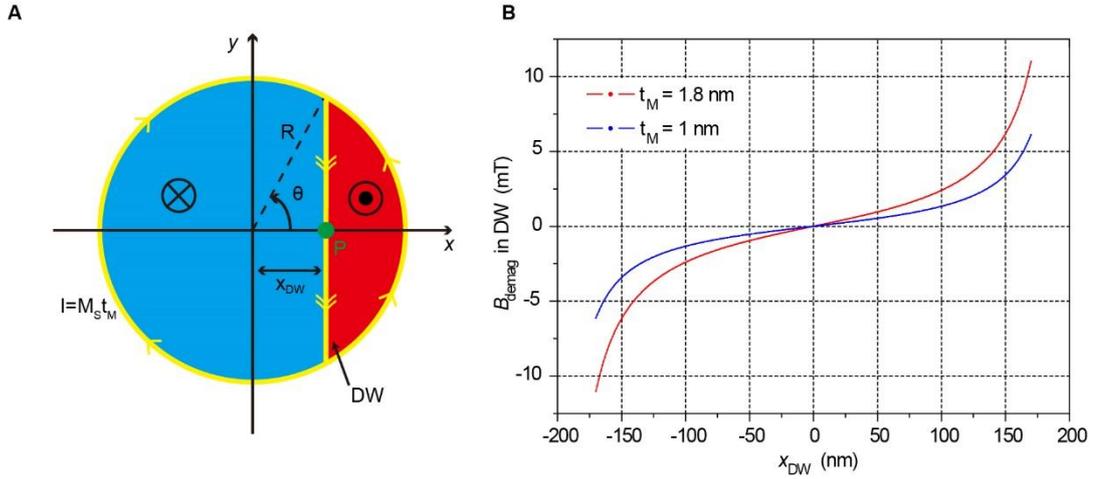

*Fig. S6. Calculation of the demagnetizing field $B_{demag}$ produced by an FL disc and acting on a DW in the disc. (A) Sketch showing the configuration assumed for the calculation and the effective magnetization current (yellow line). (B) The demagnetizing field acting on the DW as a function of the DW position $x_{DW}$ in the FL disk for different FL thicknesses $t_M$.*

## 4. Movie showing the DW pinning effect induced by the chiral vortex

We have simulated the pinning effect induced by the chiral vortex formed under the antiferromagnetic coupling of the two FLs and the opposing DMIs via OOMMF code. The simulated area, view from the +z direction, was 40 nm×60 nm. The film consists of three layers, i.e., the LwFL(ferromagnetic, 1 nm)\Spacer (nonferromagnetic, 0.2 nm)\UpFL(ferromagnetic, 1 nm). The two FLs were ferromagnetically coupled because of RRKY interactions, with a coupling strength $J_{ex}$=0.6 mJ/m$^2$. However, the two FLs lost the interlayer coupling in a 8-nm wide band area in the middle of the film. Parameters for the simulations were: $M_S$=1×10$^6$ A/m, $K_u$=1×10$^6$ J/m$^3$, $A_{ex}$=1×10$^{-11}$ J/m for each FLs. DMIs with a positive (negative) sign was set in the LwFL (UpFL), |D|=0.5 mJ/m$^2$.

In the beginning, a DW was set in the left of the film as the initial state. Then a 10 mT perpendicular field was applied continuously. We observed that the DW moved to the right and then entered the area where RKKY interactions lost. At the same time, the structure of the DW changes from Bloch type to chiral vortex. As calculated before, the chiral vortex wall can minimize both the DMI energy and demagnetizing energy, therefore, this structure served as an energy well and robustly pinned the DW. No further DW motion occurred in following. The magnetic state of the film was saved during the simulation, two frames of them can be found in Fig. S7 and the video can be found in the Supplementary material 2.

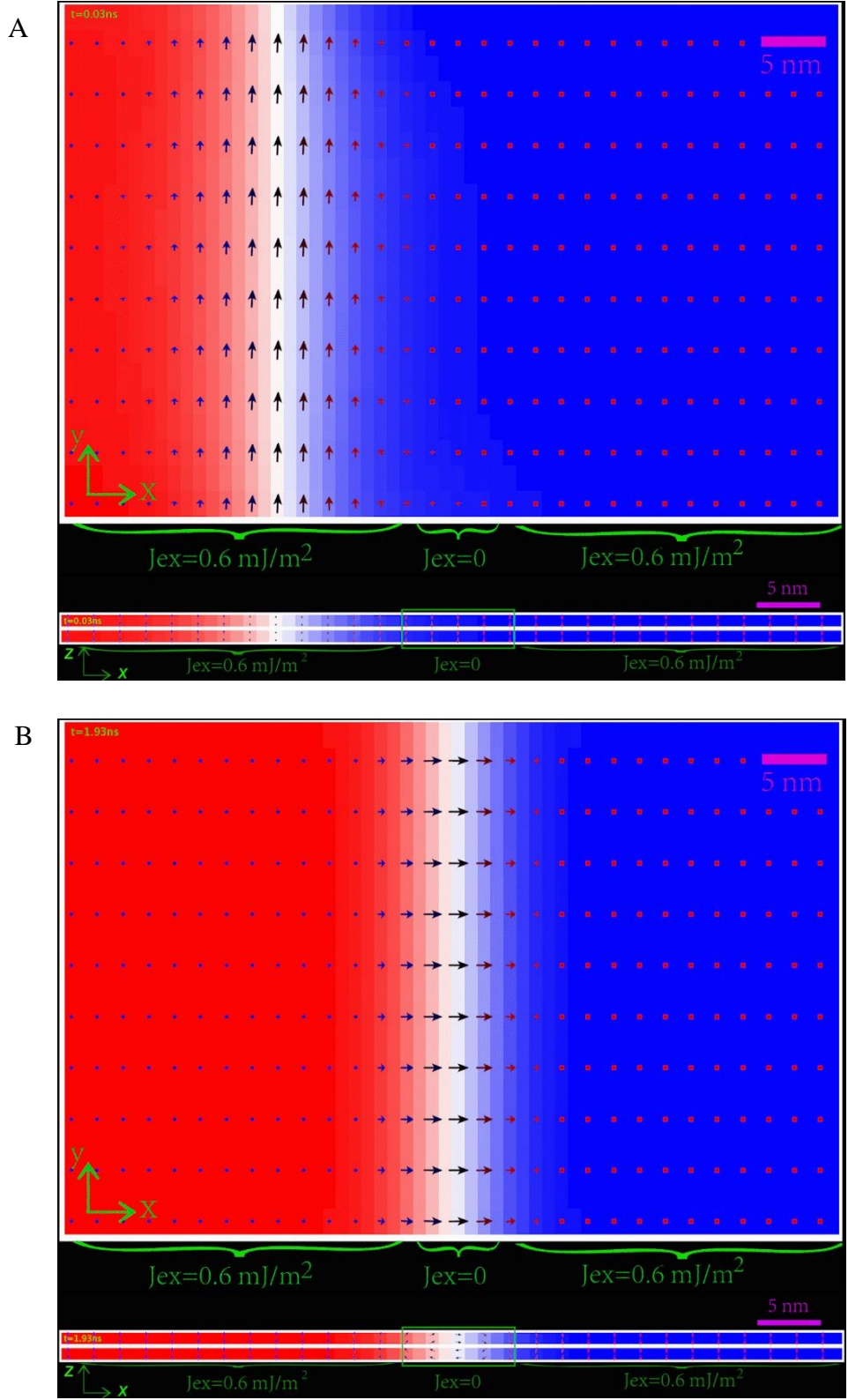

*Fig. S7. Two frames of video showing the DW pinning effect around the chiral vortex.* (**A**) *DW structure in area where the UpFL and LwFL is ferromagnetically coupled.* (**B**) *DW structure after entering an area where the coupling of the two FLs drops to zero. The video can be found in the supplementary material II. The figure in the up of each figure is the top view and figure in the lower is the side view.*